\documentclass[overfull]{epl}

\newcommand{\mc}{\mathcal}
\newcommand{\be}{\begin{equation}}
\newcommand{\ee}{\end{equation}}
\newcommand{\di}{\displaystyle}

\usepackage{bbold,psfrag}
\usepackage{color,graphics}

\def\verylongrightarrow{\relbar\joinrel\relbar\joinrel\relbar\joinrel\rightarrow}

\title{
Effective forces between colloids at interfaces
induced by capillary wave--like fluctuations}
\shorttitle{
Effective forces between colloids
}

\author{H. Lehle \and M. Oettel 
\and S. Dietrich}

\institute{                    
  Max--Planck--Institut f\"ur Metallforschung, Heisenbergstr. 3, D-70569 Stuttgart,
  and
  Institut\- f\"ur Theoretische und Angewandte Physik, Universit\"at Stuttgart,
 Pfaffenwaldring 57, D-70569 Stuttgart, Germany
}

\begin{document}

\maketitle

\begin{abstract}
We calculate the effective force mediated by thermally
excited capillary waves between spherical or disklike
colloids trapped at a fluid interface. This Casimir type
interaction is shown to depend sensitively on the
boundary conditions imposed at the three-phase contact line.
For large distances between the colloids an unexpected cancellation of 
attractive and repulsive contributions is observed
leading to a fluctuation force which decays algebraically very rapidly.
For small separations the resulting force
is rather strong and it
may play an important role in two--dimensional colloid aggregation
if direct {\it van der Waals} forces are weak.
\end{abstract}

\section{Introduction and synopsis}

The effective forces between rigid objects immersed
in a fluctuating medium have attracted a steadily growing interest
because their understanding allows one to design and tune them by choosing
suitable media and boundary conditions and by varying the thermodynamic state
of the medium. Possible applications range from micromechanical systems to 
colloidal suspensions and embedded biological macromolecules.
Accordingly, these fluctuations may be the
 zero--temperature, long--ranged quantum fluctuations of the
electromagnetic fields giving rise to the
original Casimir effect \cite{Bor01} between flat or corrugated immersed
metallic bodies \cite{Jaf04, Bue04}.
Other examples for fluctuation induced long--ranged effective forces between
immersed objects involve media such as 
bulk fluids near their critical point \cite{Kre94}, membranes \cite{Kar99} or 
interfaces 
\cite{Kai05}.

In this work we investigate the latter manifestation of  
thermal Casimir forces for nanocolloids floating
at surface--tension dominated 
liquid-vapor or liquid-liquid interfaces where they are effectively trapped. 
In the presence of charges their mutual interactions 
often exhibit unexpected strong and long-ranged attractions
\cite{Che05} inducing mesoscale pattern formation. However, these unusual patterns, which, e.g., 
are of interest for optical applications once they are fixated on
a solid substrate, appear also for neutral nanocolloids \cite{Sea99}. 
In spite of some progress ~\cite{Oet05},
the nature of these
effective forces between the colloids is not yet fully understood.
This pertains in particular to the role of fluctuations for these
types of observations. 

Therefore the present effort aims at determining the fluctuation induced
contribution to these forces generated by the inevitable thermally excited capillary
waves of the fluid interface.
As a first step the colloids are taken to be
 spherical or disklike
and electroneutral. We shall pay special attention
to the fact that these colloids are of finite size and that, through their surface
properties, they
exert certain restrictions on the three--phase contact line formed at the
intersection of
the fluid interface with the colloid surface. 

For the idealized
situation
of a pinned contact line (corresponding to Dirichlet boundary conditions) this problem
has already been studied, mostly in the context of
membrane--type interfaces
\cite{Gol96}; 
some general results can be inferred also from Ref.~\cite{Gol00}.
For a pinned contact line and a Gaussian weight for the interfacial height
fluctuations these studies yield for the asymptotic decay of the
fluctuation induced
force $F(d)$ as function of the center-to-center distance $d$ between the colloids
 $(i)$ $F(d\to \infty)\propto d^{-1}$ (attractive) if the centers of the
 colloids are 
fixed by some external means
and $(ii)$ $F(d\to \infty)\propto d^{-5}$  (attractive) if the colloids are free 
to fluctuate vertically. Going beyond this analysis we find that 
$F(d\to \infty)\propto d^{-9}$ (attractive) if, corresponding to the generic situation, 
the fluctuations of the three--phase contact line
are included, governed by a boundary Hamiltonian which is derived from the surface
properties
of the colloids and of the interface. This highlights the importance of boundary
conditions
in Casimir problems. In our case they render the long--ranged tail of the effective
force
virtually unimportant. 
However, if the colloids approach each other with a 
surface--to--surface distance $h\to 0$ we find that the force increases $\propto h^{-3/2}$.
Thus for actual systems the fluctuation induced force is effectively short--ranged  and
therefore it may play an important role in the coagulation of interfacially trapped colloids. 
This behaviour cannot be captured if the colloid is approximated by an essentially
pointlike object as it has been done in Ref.~\cite{Kai05}.
We shall also discuss experimental possibilities to
separate the fluctuation induced forces from the ubiquitous dispersion forces 
acting in colloidal systems.

\section{Model}
We consider an interface between two fluid phases I and II
at which two nano-- or microscopic colloids are trapped,
either spherical with radius $R$ or disklike with radius $R$ and thickness $H$.
Since the weight of such colloids  is negligible, the equilibrium (or 
reference) configuration  is the flat interface with the centers of the 
spherical colloids vertically positioned such that Young's law holds at the
horizontal 
three-phase contact circle with radius $r_0 = R \sin\theta$.
Young's angle
is determined by
$\cos\theta =(\gamma_{\rm I}-\gamma_{\rm II})/\gamma$ where $\gamma$ is the surface
tension
of the I/II interface, and $\gamma_{{\rm I}\,[{\rm II}]}$  the surface
tension between
the colloid and phase I [II]. For the disks, the contact line is either the
upper ($\theta<\pi/2$) or lower ($\theta>\pi/2$) circular edge, so that $r_0 = R$.
Deviations from the planar reference interface $z=0$ are assumed to be small
which allows us to
use the Monge representation $(x,y,z=u(x,y))=({\bf x},z=u({\bf x}))$ as a
parametrization of the actual interface positions.
The free energy cost for thermal fluctuations 
with small gradients around the flat interface
is determined by the the change in interfacial 
energy of {\em all} interfaces (I/II, colloid/I [II]):
\begin{equation}\label{ham}
\mc{H}_{\rm tot}=\mc{H'}_{\rm cw}
+\gamma_{\rm I}\Delta A_{\rm I}
+\gamma_{\rm II}\Delta A_{\rm II}
=\frac{\gamma}{2} \int_S d^2x\, \left[ (\nabla u)^2
+\frac{u^2}{\lambda_{c}^{2}} \right] 
+\gamma_{\rm I}\Delta A_{\rm I}
+\gamma_{\rm II}\Delta A_{\rm II}
\; .
\end{equation}
Here, $\mc{H'}_{\rm cw}$ is the standard capillary--wave
Hamiltonian. 
In eq.~(\ref{ham}) the capillary length is given by 
$\lambda_c = [\gamma/(|\rho_{\rm II}-\rho_{\rm I} |\, g)]^{1/2}$, where $\rho_k$
is the mass density in phase $k$ and $g$ is the gravitational constant. 
The first term in $\mc{H'}_{\rm cw}$ expresses the energy needed for
creating the additional interface area associated with the
height fluctuation, the second one the corresponding cost in gravitational
energy. Usually, one has $\lambda_c \gg R$; however, care is required in taking
the limit $\lambda_c \to \infty$ (see below),
corresponding to the limit of a vanishing pinning force $g\to 0$.
 The integration domain
$S$ is given by the plane $\mathbb{R}^2$ ($z=0$) minus the enclosed areas
$S_{1}$ and $S_2$
of the projections 
of the contact lines on the first  and second colloid, respectively, onto 
this plane. Thereby the Hamiltonian
depends explicitly on the positions of the colloids.
In the reference configuration, the projections $S^0_i$ of the contact lines  
are the aformentioned circles  with boundary $\partial S^0_i$ and common
 radius $r_0$.
As discussed in the Introduction, in the case of spheres the contact line
itself may
fluctuate around its reference position 
with an energy cost which is determined by 
changes of the projected meniscus area $S$ and of the interfaces colloid/I,II 
($\Delta A_{\rm I,II}\neq 0$ in eq.~(\ref{ham}))
with respect to the reference configuration.
We introduce the vertical position 
of the contact line at colloid $i$ as a function of a polar angle $\varphi_i$ defined
on $\partial S^0_i$
by
$f_i\equiv u(\partial S^0_i)=P_{i0}+\sum_{m\ge 1} [ P_{im}\cos m\varphi_i +Q_{im}\sin
m\varphi_i ]$.
The expansion coefficients $P_{im}$ and $Q_{im}$ are referred to as multipole
moments of 
the contact line. Furthermore, $h_i$ is introduced as the fluctuation induced 
change in vertical
position of the
colloid centers with respect to the reference configuration. 
%
Following Ref.~\cite{Oet05a}, we restrict the integration in 
$\mathcal{H'}_{\rm cw}$ 
to the reference domain $\mathbb{R}^2\setminus \bigcup
S^0_i$ and expand 
the remaining energy differences up to second order in $f_i$ and $h_i$, 
introducing a boundary 
Hamiltonian $\mathcal{H}_{{\rm b}, i}$ so that \nolinebreak
$\mc{H}_{\rm tot}=\mc{H}_{\rm cw}+\sum_i\mc{H}_{\rm b,i}$:
\begin{eqnarray}
\mathcal{H}_{\rm cw}[ u]
&=&\frac{\gamma}{2} \int_{\mathbb{R}^2\setminus \bigcup S^0_i} d^2x\, \left[
(\nabla u)^2
+\frac{u^2}{\lambda_{c}^{2}} \right] 
\; , \\
\label{Hb}
\mc{H}_{{\rm b},i}[f_i,h_i]
&=&\frac{\gamma}{2 }\,\oint_{\partial S^0_i} d\varphi_i\,(f_i-h_i)^2
=\pi\gamma\left[ (P_{i0}-h_i)^2+ \frac{1}{2}\sum_{m\ge 1} \left(
P_{im}^2+Q_{im}^2\right)\right]\;.
\end{eqnarray}
The effective force $F(d)=-\frac{\partial \mc{F}}{\partial d}$ 
as function of the mean local distance
between the colloid centers is determined by 
 the free energy
${\mathcal F}(d)= -k_{\rm B} T (\ln {\mathcal Z}(d))$. The partition function
${\mathcal Z}(d)$ is obtained by a functional integral over all possible interface 
configurations $u$ and $f_i$; the boundary configurations are included by 
$\delta$-function constraints, as introduced in Ref.~\cite{Li91}: 
\begin{equation}\label{Z1}
\mathcal{Z}=\mathcal{Z}_0^{-1} \int \mathcal{D}u\,
\exp \left\{-\frac{\mathcal{H}_{\rm cw}[u] }{k_{\rm B}T} \right\}\;
\prod_{\rm i=1}^2 
\int \mathcal{D}f_i
\prod_{{\bf x}_i \in \partial S^0_{i}}
\delta [u({\bf x}_{ i})-f_{ i}({\bf x}_{ i})]
\exp \left\{-\frac{\mathcal{H}_{{\rm b},i}[f_i,h_i]}{k_{\rm B}T} \right\}
\;.
\end{equation}
$\mathcal{Z}_0$ is a normalization factor such that $\mathcal{Z}(d\to\infty)=1$.
We shall discuss two different realizations of the boundary conditions for
the contact line 
(see Fig. \ref{fig1}).
 (A) The contact lines and the vertical colloid positions fluctuate
freely;
this corresponds to the physical situation of smooth, spherical colloids.
In this case, the integration measure is given by $\mc{D}f_i=  \int dh_i\prod_m dP_{
im}dQ_{im}$.
(B) The contact lines are pinned in the reference configuration 
and do not fluctuate with respect to the colloid surfaces.
This corresponds to disks or Janus spheres consisting of two different materials.
Within the pinning case, we furthermore distinguish the following 
three 
physical
situations:
(B1) The colloid positions are fixed, 
e.g., by using optical tweezers;
 thus there are no
integrations over the boundary terms. (B2) The vertical positions of the colloids 
fluctuate freely, so that $\mc{D}f_i= dh_i dP_{i0}\delta(P_{i0}-h_i)$.
(B3) The vertical position and the orientation of the colloids (tilts) 
fluctuate freely. Up to second order in the tilts this corresponds
to $\mc{D}f_i=dh_i dP_{i0} dP_{i1}dQ_{i1}\delta(P_{i0}-h_i)$. The $\delta$-function
expresses the pinning condition.
All these cases can be discussed conveniently after splitting the
field $u$ of the local interface position into 
a mean--field and a fluctuation part, $u=u_{\rm mf}+v$. The
mean--field part solves the Euler--Lagrange equation
$ (-\Delta+\lambda_c^{-2})\,u_{\rm mf}=0$ with the boundary condition
$u_{\rm mf}\,|_{ \partial S^0_{i}}=f_{ i} $. Consequently the fluctuation part
vanishes at the contact line:
$v\,|_{ \partial S_{ i}^0}=0$.
Then the partition function  $\mathcal{Z}=\mathcal{Z}_{\rm fluc}\mathcal{Z}_{\rm mf}$ 
factorises 
-- due to the Gaussian form of $\mc{H}_{\rm cw}$ --
into a product of a fluctuation part 
independent of
the boundary conditions 
 and a  mean field part 
depending on the boundary conditions (which may fluctuate themselves, see the cases 
(A), (B2), and (B3)): 
%
\begin{eqnarray}\label{Zsep}
 \mathcal{Z}_{\rm fluc} &=& \mathcal{Z}_0^{-1} \int \mathcal{D}v\,
\prod_{\rm i=1}^2 \prod_{{\bf x}_i \in \partial S^0_i}\delta ( v({\bf x}_i))
\exp \left\{-\frac{\mathcal{H}_{\rm cw}[v]}{k_{\rm B}T} \right\}\;,  \nonumber \\
 \mathcal{Z}_{\rm mf} &=& \prod_{\rm i=1}^2
\int \mathcal{D}f_i
\, \exp\left\{ -\frac{\gamma }{2 k_{\rm B}T} \sum_{i}
\oint_{\partial S^0_i}d\ell_i f_i({\bf x_i})\,( \partial_n u_{\rm mf}(\bf{x}_i))\right\}
\exp \left\{-\frac{\mathcal{H}_{{\rm b},i}[f_i,h_i]}{k_{\rm B}T} \right\}
 \;.
\end{eqnarray}
The first exponential in $\mathcal{Z}_{\rm mf} $ stems from applying Gauss' theorem
to the capillary wave Hamiltonian $\mc{H}_{\rm cw}[u_{\rm mf}+v]$.
In this term
$ \partial_n u_{\rm mf}$ denotes the normal derivative of the mean--field
solution towards the interior of the circle $\partial S^0_i$, and $d\ell_i$ is
the infinitesimal line segment on $\partial S^0_i$. 
In the following we provide only the main steps of the
subsequent evaluation of both the fluctuation and the mean-field part. For more
details the reader is referred to Ref.~\cite{Leh06}.

\section{Fluctuation part}
The fluctuation part appears in all cases introduced above. 
In the case
(B1) it constitutes the full result for the partition function because 
in that case $u_{\rm mf}=0$
and $\mc{Z}_{\rm mf}=1$.
The $\delta$-functions 
in the fluctuation part of the partition function
can be removed by using their integral representation via
 auxiliary fields $\psi_i ({\bf x}_i)$ defined on the
interface boundaries $\partial S^0_i$ ~\cite{Li91}. This enables us to integrate out 
the field $u$ leading to
\begin{eqnarray}
\label{aux}
 \mathcal{Z}_{\rm fluc} =
\int \prod_{i=1}^2 \mc{D}\psi_i\,
\exp\left\{
-\frac{k_{\rm B}T}{2\gamma}\sum_{i,j=1}^2
\oint_{\partial S^0_i}d\ell_i \oint_{\partial S^0_j}d\ell_j\,
\psi_i ({\bf x}_i)\,G(|{\bf x}_i-{\bf x}_j|)\,\psi_j({\bf x}_j)\right\}\;.
\end{eqnarray} 
Here, we introduced Green's function 
$G({\bf x})=K_0(|{\bf x}|/\lambda_{  c})/(2\pi)$
of the operator $(-\Delta + \lambda_c^{-2})$
where $K_0$ is the modified Bessel
function of the second kind.
In this form, the fluctuation part  resembles 2d screened electrostatics: it is the 
partition function of a system
of fluctuating charge densities $\psi_i$ residing on the contact circles. For large
$d/r_0$ it can be calculated by a multipole expansion 
~\cite{Gol96}.
In this limit and for $\lambda_c/d \to \infty$ 
we find for the fluctuation force:
\begin{equation}
\label{ffluc}
F_{\rm fluc}=k_{\rm B}T \frac{\partial}{\partial d}\ln \mathcal{Z}_{\rm fluc}
  \;\to\;-\frac{k_{\rm B}T}{2}\frac{1}{d\ln (d/r_0)} + \mc{O}(d^{-3})\;,
\qquad \frac{d}{r_0}\gg 1,\;\frac{d}{\lambda_c}\to 0\;.
\end{equation}
Note, however, that here the limit $\lambda_c \to \infty$
is attained slowly  with a leading correction term of the order
$1/(d\ln(\lambda_c/r_0))$, 
and that in this limit the free energy difference 
${\cal F}(d)-{\cal F}(\infty)$ is actually 
ill--defined (${\cal F}(d) \sim \ln\ln (d/r_0)$)
and therefore the effective colloidal interaction in case (B1) -- fixed
colloids and
pinned interface -- is only meaningful for a finite capillary length
$\lambda_c$
\cite{comment1},
 similar to  a free two--dimensional interface the width
 of which is determined by the 
 capillary wave fluctuations and diverges logarithmically $\sim\lambda_c$.

\begin{figure}
 \begin{minipage}{\textwidth}
\psfrag{2}{ $2$}
\psfrag{5}{ $5$}
\psfrag{10}{ $10$}
\psfrag{15}{ $15$}
\psfrag{20}{ $20$}
\psfrag{25}{ $25$}
\psfrag{2.4}{ $2.4$}
\psfrag{2.8}{ $2.8$}
\psfrag{1e02}{ $100$}
\psfrag{1e01}{ $10$}
\psfrag{1e0}{ $1$}
\psfrag{1e-02}{ $10^{-2}$}
\psfrag{1e-04}{$10^{-4}$}
\psfrag{1e-06}{$10^{-6}$}
\psfrag{Ffluc}{(B1)}
\psfrag{Fmfmon}{(B2)}
\psfrag{Fmfdip}{(B3)}
\psfrag{casea}{}
\psfrag{casec1}{$-F_{\rm fluc}$}
\psfrag{caseb2}{\begin{minipage}{4cm}
\vspace*{0.5cm}\hspace*{-0.4cm}$\big\}\;\, F_{\rm mf}$
$ \begin{array}{l} \mbox{(up to monopoles)}\\ 
 \mbox{(up to dipoles)}\end{array}$
\end{minipage}}
\psfrag{b}{\bf{(a)}}
 \psfrag{d}{\large $d/R$}
 \psfrag{F}{\large $F/(k_{\rm B}T/R)$}
 \onefigure[width=0.98\textwidth]{mfflucplot.eps}
   \psfrag{I}{\footnotesize I}
   \psfrag{II}{\footnotesize II}
   \psfrag{d}{\footnotesize $d$}
   \psfrag{R}{\footnotesize $R$}
   \psfrag{h}{\footnotesize $h$}
   \psfrag{q}{\footnotesize $\theta$}
   \psfrag{z=0}{\footnotesize $z=0$}
 \vspace*{-2.5cm}
 \hspace*{-0.18\textwidth} 
\onefigure[width=0.39\textwidth]{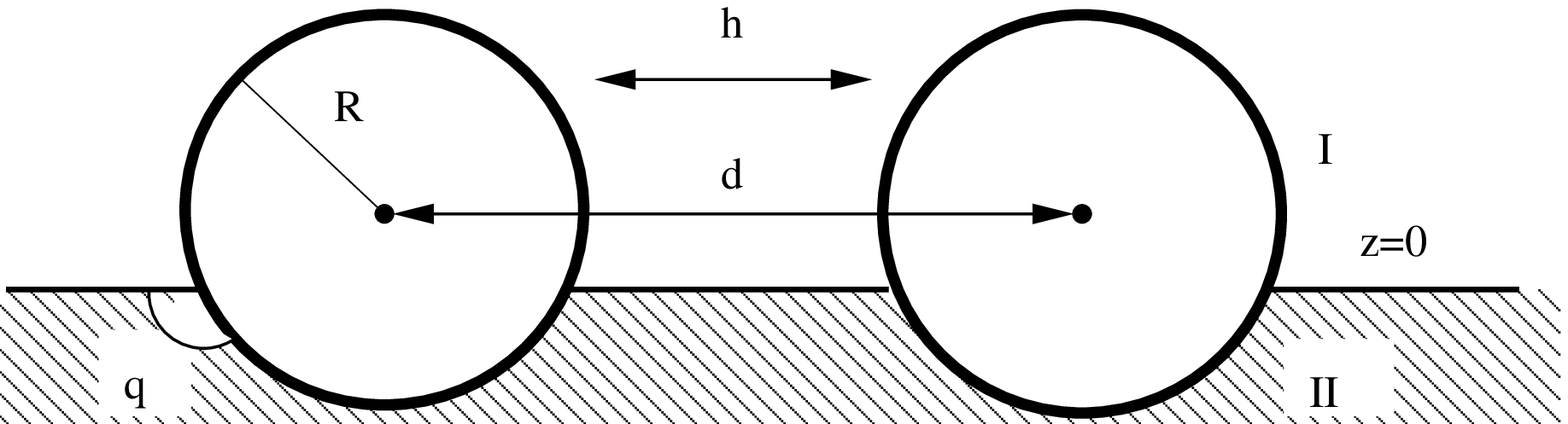}
 \vspace*{1.3cm}
 \end{minipage}
\vspace*{0.8cm}

{
\psfrag{a}{\bf{(b)}}
\psfrag{A}{A}
\psfrag{B1}{B1}
\psfrag{fixed}{fixed}
\psfrag{B2}{B2}
\psfrag{B3}{B3}
\oneimage[width=\textwidth]{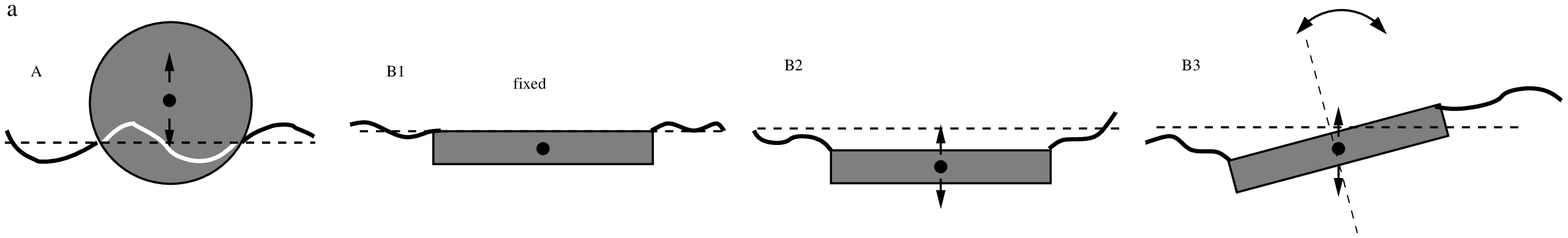}
}
\caption{
(a) Numerically calculated thermal total Casimir force $F$ (symbols) 
for various boundary conditions
compared with the analytic results in eq. (\ref{lrforces}) (lines).
The inset shows the results  for the two contributions to the short range regime:
numerically calculated
fluctuation contribution ($-F_{\rm fluc}$) and
     mean-field contribution  ($F_{\rm mf}$)  
   compared with the analytic expressions in eqs. (\ref{ffluc2})
    (dashed) and (\ref{derjmf}) (dotted), respectively.
(b) Sketch of the various boundary conditions. 
For (B1)-(B3), the disks may be replaced by Janus spheres.
}
\label{fig1}
\end{figure}

 In the
opposite limit of small surface--to--surface distance $h=d-2r_0$ the
fluctuation force can be calculated by using the 
well--known result for the fluctuation force per length $f_{\rm 2d}(\tilde h)=-
k_{\rm B}T\,\pi/(12 {\tilde h}^2)$ 
between two lines
a mean distance $\tilde h$ apart
\cite{Li91}, together with the Derjaguin (or proximity) approximation \cite{Der34}:
%
\begin{eqnarray}
\label{ffluc2}
F_{\rm fluc} \approx  
-\frac{\pi k_{\rm
B}T}{12}\int_0^{r_0}dy\,\frac{1}{\left(h+2r_0-2\sqrt{r_0^2-y^2}\right)^2}
&\stackrel{r_0/h \to \infty}{ \verylongrightarrow}&
-k_{\rm B}T\, \frac{\pi^2}{48}\,
\frac{r_0^{1/2}}{h^{3/2}}
+\mc{O}(h^{-1/2})
 \;.
\end{eqnarray}
This strong increase as $h\to 0$ is a consequence of the finite
(mesoscopic) size of the colloids and is missed if the colloids are approximated as
pointlike
objects
\cite{Kai05}. It also turns out (see below) that this increase is not
 cancelled by any mean--field contributions. Thus it constitutes the dominant
short--ranged contribution in all cases (A) and (B1)--(B3). 
For intermediate distances $d$ between these limits  $F_{\rm fluc}$ 
must be evaluated numerically. In eq.~(\ref{aux}) the integral over 
 the auxiliary fields $\psi_i$ can be carried out because they appear only 
quadratically in the exponent. The resulting 
determinant is divergent and requires regularisation. However, 
the derivative of its logarithm with respect to $d$ 
(corresponding to the force) is finite and convergent in a numerical analysis
(see Refs.~\cite{Bue04,Leh06} for further details). Actual numerical results,
which recover the limiting behaviors given by eqs. (\ref{ffluc}) and (\ref{ffluc2}),
 are presented in Fig.\nobreak \ref{fig1}.

\section{Mean--field part}

The calculation of  $\mc{Z}_{\rm mf}$ (Eq.~(\ref{Zsep})) requires to determine
the solution of the differential equation $(-\Delta+\lambda_c^{-2})\,u_{\rm mf}=0$
for the (fluctuating) boundary conditions $u_{\rm mf}({\bf x}_i)=f_i({\bf x}_i)$
for ${\bf x}_i \in \partial S^0_i$ 
and $u_{\rm mf}({\bf x})\to 0$ 
for $|{\bf x}|\to \infty$.
We 
use the superposition ansatz $u_{\rm mf} = u_1+u_2$ 
where $u_i= \sum_m  K_m ( r_i/\lambda_{ c})\left[
A_{im} \cos m\varphi_i + B_{im}\sin m \varphi_i
\right]$ is 
the general mean--field solution in $\mathbb{R}^2\setminus  S^0_i$. 
The solution has to match to the boundary conditions 
at both circles $\partial S^0_1$ and $\partial S^0_2$.
This can be achieved by  a projection of $u_2$ onto 
the complete set of functions on $\partial S^0_1$, $\{\cos m\varphi_1,\sin
m\varphi_1\}$, 
and vice versa. Equating this expansion with the 
multipole moments ${\bf \hat{f_i}}=(P_{i0},P_{i1},Q_{i1},\dots )$
of the values $f_i({\bf x}_i)$ of the field $u(\bf x)$
at the contact lines 
leads to a system of linear equations for the expansion coefficients
$\{A_{im},B_{im}\}$.
This system can be solved analytically within a systematic $1/d$ expansion or
numerically, 
observing rapid
convergence (even for small $d$), allowing us to truncate the
expansions at $m_{\rm max}\approx 20$.
The mean field part of the partition function 
can then be
written as
\begin{equation}
\mc{Z}_{\rm mf} = \int \mc{D} f_i\; \exp\left \{ - \frac{\mc{H}[u_{\rm mf}]}{k_{\rm
B} T}\right\}\;
  \exp\left\{ -\frac{\pi\gamma}{k_{\rm B} T}\sum_i (P_{i0}-h_i)^2   \right\},
\end{equation}
where $\mc{H}[u_{\rm mf}]$ is a symmetric quadratic form of the multipole moments of the
contact lines, 
 \begin{eqnarray}\label{Hmean}
 \mc{H}[u_{\rm mf}]
&=&
 \frac{\gamma}{2}
 \left(
 \begin{array}{c}
 {\bf\hat{f}}_1\\
 {\bf\hat{f}}_2
 \end{array}
 \right)
 ^{\rm T}
 \left(\begin{array}{cc}
 {\bf E}_{\rm 1\,self} & {\bf E}_{\rm int} \\
 {\bf E}_{\rm int} & {\bf E}_{\rm 2\, self}
 \end{array}\right)
 \left(
 \begin{array}{c}
 {\bf \hat{f}}_1\\
 {\bf \hat{f}}_2
 \end{array}
 \right)\;,
 \end{eqnarray}
so that the $d$--dependent part of $\mc{Z}_{\rm mf}$ is given by 
$\det {\bf E}$.
Before discussing the analytic structure of the dependence on $d$, we recall that
the integration measure $\mc{D} f_i$ differs for the cases
(A)
, (B2) 
, and (B3)
. In all cases,
 the large $d$ expansion of the mean--field part
of the partition function 
(with $\lambda_c\to \infty$) leads to a repulsive effective force between the colloids:
\begin{equation}
\label{fmf}
F_{\rm mf}=k_{\rm B}T \frac{\partial}{\partial d}\ln \mathcal{Z}_{\rm mf}
  \to \frac{k_{\rm B}T}{2}\frac{1}{d\ln (d/r_0)} +
  \mc{O}(d^{-3})\;,\qquad\frac{d}{r_0}\gg 1\;,\frac{d}{\lambda_c}\to 0\;.
\end{equation}
Due to $\mc{Z}=\mc{Z}_{\rm fluc}\mc{Z}_{mf}$
the total effective force 
 is  $F=F_{\rm fluc}+F_{\rm mf}$.
 The leading terms in $F_{\rm mf}$ and
$F_{\rm fluc}$ (Eqs.~(\ref{fmf}) and (\ref{ffluc})) cancel.
The same holds for the first subleading terms, and in the cases
(A) and (B3) also for the second-next subleading terms. 
Due to these cancellations it turns out that in all four cases
for large distances the effective force is attractive with 
 the leading term stems from the fluctuation part.
This 
can 
 be
 summarized as follows ($\frac{d}{r_0}\gg 1\;,\frac{d}{\lambda_c}\to 0$): 
 \begin{eqnarray*}
 \label{lrforces}
  \begin{array}{lll}
   F \to {\di-\frac{\di 8a k_{\rm B}T}{\di r_0}\,\left(\frac{\di r_0}{\di
 d}\right)^9 }
& \hspace*{-0.4cm}\bigg\{\!
\begin{array}{l}
\mbox{(A)}  \\
 \mbox{(B3)} 
\end{array}
&\!\!\!
\begin{array}{l}
 \mbox{no pinning -- fluctuating contact line: $a=1$}   \\
\mbox{pinning -- colloidal height and tilt
 fluctuations: $a=9$}
\end{array}\vspace*{0.1cm}
    \\
    F \to {\di-\frac{\di 4 k_{\rm B}T}{\di r_0}\,\left(\frac{\di r_0}{\di d}\right)^5}\;,
  &
   \mbox{(B2)} 
& 
\mbox{pinning$\,$ --$\,$ colloidal$\,$ height$\,$ fluctuations} \\
    \vspace*{-0.25cm}
\\
   F \to {\di - \frac{\di k_{\rm B}T}{\di 2}\frac{\di 1}{\di d\ln (d/r_0)}} \;,
  & \mbox{(B1)} & \mbox{pinning$\,$ --$\,$fixed $\,$ colloids.}
\hspace*{4.45cm}(\theequation) \addtocounter{equation}{1}  
\vspace*{0.1cm}
  \end{array}
 \end{eqnarray*}
 In the opposite limit, $h=d-2R \to 0$ the effective mean--field force $F_{\rm mf}$
 is repulsive. In case (B2) the leading behaviour for $h\to 0$ can be estimated
 analytically ~\cite{Leh06}:
 \begin{equation}\label{derjmf}
  F_{\rm mf}(h\to 0) \approx \frac{k_{\rm B}T}{4\, h}
 \end{equation}
   We find numerically that the inclusion of higher (beyond the zeroth order) multipole moments
   of the contact line, as relevant for the cases (A) and (B3), slightly changes the
   prefactor
   in Eq.  (\ref{derjmf}) but does not alter the $1/h$ behaviour. 
   The  increase of the mean--field force for $h\to 0$
is weaker
than that of $F_{\rm fluc}$ 
   so that for all cases the leading behaviour of the total effective force
   at small distances is also attractive and given by Eq.~(\ref{ffluc2}).


\section{
Discussion} 
The restrictions which 
two rotational symmetric colloids
trapped at a fluid interface impose 
on its thermally excited 
capillary waves,
 lead to a thermal Casimir force between them.
At large separations $d\gg r_0$ 
there is an interesting interplay between
the attractive interaction due to the interface fluctuations
and a repulsive interaction generated by the 
mean-field contribution to the
fluctuating
boundary conditions.
This results in a cancellation of leading terms
up to a certain order in $1/d$, which depends on
the specific 
type of
boundary conditions.  
In the opposite limit of small 
separations $h=d-2r_0 \ll r_0$,
the effect of the boundary conditions is much less pronounced, 
and the resulting  force is dominated by $F_{\rm fluc} \sim h^{-3/2}$ (attractive)
 compared with $F_{\rm mf}\sim h^{-1}$ (repulsive), leading to a strong
 attractive Casimir
 interaction in this regime (see Fig. \ref{fig1}).
Note 
that for  colloid--colloid separations $h$  of the order
of the molecular length scale $\sigma$ of the fluids, the capillary--wave model
is no longer valid.  
In this limit
the total effective force stays finite and can
be understood by 
taking into account the formation of a
wetting film around the 
colloids \cite{baubieksd00}.
At small separations typically also  {\it van der Waals} (vdW)  forces become
important, leading to a strong tendency of colloid aggregation if
not compensated by a repulsive interaction.
For spherical  
colloids at small separations and $\theta=\pi/2$ the well--known Derjaguin result
for the vdW force
is given by 
$F^{\rm vdW}_{\rm Derj} = -A_{\rm H}/(12\pi)\,R/h^2$
where the (effective) Hamaker constant $A_{\rm H}$ is determined
by the frequency dependent dielectric permittivities of
both the colloids and the two fluid phases. $F^{\rm vdW}_{\rm Derj}$
shows an even stronger increase for $h\to 0$
and hence will dominate the Casimir force (\ref{ffluc2}) since 
the Hamaker constant is typically $A_{\rm H}\approx 1 \ldots 10 k_{\rm B}T$. 
However, refractive index matching
between colloids and fluids can result in 
a significantly smaller value of 
$A_{\rm H}$.
For $A_{\rm H}/k_{\rm B}T < (\pi^3/4)\sqrt{h/R}$
we have $|F_{\rm fluc}| > |F^{\rm vdW}_{\rm Derj}|$
and therefore in this distance regime an increased influence of the 
fluctuation induced force on particle aggregation.
In the case of two cylindrical disks with height $H$ the Derjaguin
approximation leads to 
$F^{\rm vdW}_{\rm Derj} 
\sim -(H/h^2)\,(R/h)^{1/2}$ 
if $ h\ll R,\,H$.
However, for very {\it t}hin {\it d}isks ($H\ll h\ll R$,
c.f. cases (B1)--(B3)),
integration over atomic pair potentials $\sim r^{-6}$,
which give rise to 
 the vdW force, results in
\begin{equation}
  \label{vdWd2}
  F^{\rm vdW}_{\rm td}
  =
  - A_{\rm H}\frac{15 \pi^2}{48}\frac{H^2}{h^2}\sqrt{\frac{R}{h}}\frac{1}{h}\;.
  \end{equation}
For these systems $F_{\rm vdW}/F_{\rm fluc}=15 (A_{\rm H}/k_{\rm B}T)(H/h)^2\ll1$,
so that for $H\ll h \ll R$ the  driving mechanism for flocculation 
 is entirely given by the fluctuation induced force.
Thus, the fluctuation induced force can strongly enhance the
tendency of colloids at interfaces to flocculate. This effect is 
independent of material parameters, as long as the system is in the
capillary wave regime, i.e., for $h\gg \sigma$.
In order to discriminate the fluctuation induced force from the vdW force we propose
three experimental scenarios. The first one is to 
weaken the vdW force by  lowering the  Hamaker constant via 
index matching which increases the importance
of the fluctuation force.
Second, for thin disks, the fluctuation induced force dominates, 
independent of the dielectric properties of the materials. 
 A third mechanism consists of reducing the vdW force by a light polymer coating
 of the colloids so that in eq.  (\ref{ffluc2})  $h=h_{\rm sts}$  
 is given by the {\it s}urface--{\it t}o--{\it s}urface
 distance between the polymer shells, whereas the distance $h=h_{\rm core}>h_{\rm sts}$
between the cores of the two colloids enters into $F^{\rm vdW}_{\rm Derj}$. 
Thus for a sufficiently thick polymer shell the fluctuation induced force
will be the dominant interaction at small distances.
%
%



\begin{thebibliography}{99}
\bibitem{comment1}
{This explains why the results for the case of a two--dimensional
fluctuating medium
appear to be somewhat ill--defined in Ref.~\cite{Gol00} (see Eq.~(19) therein).
}
\end{thebibliography}
\end{document}